\documentclass[thmsa,onecolumn,a4paper]{article}
\usepackage{amssymb}
\usepackage{sw20jart}
\usepackage{amsmath}
\usepackage{graphicx}

\input{tcilatex}
\begin{document}

\author{Bertrand Chauvineau}
\title{A family of scalar-Einstein and vacuum Brans-Dicke axisymmetric
solutions}
\date{Universit\'{e} C\^{o}te d'Azur, Observatoire de la C\^{o}te d'Azur,
CNRS, Laboratoire Lagrange, France\\
(e-mail : Bertrand.Chauvineau@oca.eu)}
\maketitle

\begin{abstract}
A new method is proposed, that establishes a one to one correspondance
between the whole set of static axially symmetric vacuum GR solutions and a
specific class of stationary axially symmetric scalar-Einstein ($%
R_{ab}=\partial _{a}\varphi \partial _{b}\varphi $) solutions having a given
mass and a given angular momentum. The method explicitly takes advantage of
the Kerr metric Ricci flatness. This also results in a class of stationary
axially symmetric vacuum, ie Kerrlike, Brans-Dicke solutions. A particular
solution, that is asymptotically flat, is more closely considered. It
converges to Kerr for a vanishing scalar charge, but fails to converge to
the Fisher-Janis-Newman-Winicour solution for a vanishing "rotation
parameter". This solution exhibits a naked singularity having a ringlike
structure.
\end{abstract}

\baselineskip12truept

\bigskip

\noindent \textbf{I- Introduction}

Obtaining exact solutions of a relativistic gravity theory is far from being
an easy task, due to the highly non-linear character of the involved field
equations. The most known solutions are Schwarzschild, Kerr, and
Friedmann-Lemaitre-Robertson-Walker (FLRW), that solve the general
relativity (GR) field equation in vacuum for the two first ones. These three
solutions (or family of solutions for FLRW) are of obvious usefulness in the
astronomical framework. Some other exact GR solutions are known, that are
also very useful for astronomical purposes [1][2][3]. On the other hand, an
incredibly huge number of solutions, the usefulness of which is not obvious
a priori, have been obtained so far [4]. However, despite the non immediate
relevance of a solution, its usefulness should not be underestimated.
Indeed, any solution may reveal some unexpected property of the considered
theory. It may also happen that a solution only later turns out to be of
genuine astronomical relevance: the emergence of the black hole (BH) concept
from the Schwarzschild solution is probably the most obvious exemple.

Since many attemps to quantify gravity, and/or to unify gravity with other
interactions, return a classical gravitationnal sector having not a GR, but
a scalar-tensor (ST) structure [5][6][7], Brans-Dicke (BD) and ST gravity
theories are considered as valuable alternatives to GR, despite the fact
that the latter successfully passes solar system tests up to now [8]. (Let
us also point out that many ST theories are driven to mimic the GR behaviour
as a consequence of the cosmic expansion [9][10].) Thence the interest in ST
theories, and specifically in BD, that just involves a constant parameter $%
\omega $ instead of an arbitrary function $\omega \left( \Phi \right) $. In
this context, looking for exact BD/ST solutions is particularly appropriate
to point out relevant qualitative features with respect to GR. For instance,
the spherical Brans class I solution [11][12] generally exhibits a naked
singularity (NakS) or wormhole structure [13] (see also [14] for pionering
works on the detailled significance of the Class I, II, III and IV Brans
solutions), such features being absent from the Schwarzschild GR solution.
Besides, it has been recently shown that a particle orbiting a large $\omega 
$ Brans class I solution results in an observed (by a far observer) unbound
orbital frequency, depending on how much the solution is scalarized [15].
One may then suspect striking qualitative differences in extreme mass ratio
binaries BD/ST gravitational radiation, with respect to GR, since GR orbital
frequencies can not exceed the innermost circular orbital value.
Nevertheless, let us remind that under some conditions (mainly regularity,
asymptotical flatness and finite area horizon), vacuum and stationary BH
like solutions are the same in BD/ST as in GR [16][17].

It is known from long that, by the means of a conformal transformation, any
vacuum BD solution is associated to a massless scalar filled GR solution
[18]. Thence seeking vacuum BD solutions can be reformulated as a scalar-GR
problem. From the Hawking's theorem [16], it is clear that any stationary
axisymmetric (SAS) vacuum BD, but non Kerr, solution should exhibit a NakS
structure, unless exhibiting some peculiar feature that allow it to evade
the theorem (like being not asymptotically flat).

A method has been proposed by [19], that allows to generate a BD vacuum SAS
solution from a GR vacuum SAS seed one, provided one is able to solve a
given non linear PDE system. A class of electrovacuum BD solutions has been
obtained by [20], that generalises the Majumdar-Papapetrou GR solution. The
same authors later proposed a method that allows to generate SAS vacuum BD
solutions from SAS vacuum GR ones (and also from static axisymmetric vacuum
BD ones) [21], but the solutions then obtained are generally not
asymptotically flat. The SAS, but only one coordinate dependent, case has
been solved by [22]. The case of ST SAS solutions has been considered by
[23], and explicit solutions are given in a few case of very specific ST
(not BD) theories. Along the same lines as [19], it is claimed in [24] that
any SAS vacuum BD solution can be obtained by nonlinearly combining any SAS
vacuum GR solution and any vacuum solution of the Weyl class. A Kerr-type BD
solution has been obtained in [25], that was built in such a way that it
reduces to the Kerr solution in the $\omega \longrightarrow \infty $\ limit.
Inspired by [20] and [22] works, a BD BH-like solution is proposed by [26],
but this solution is not asymptotically flat. Motivated by particle
collisions near a Kerr-like BD BH, another non asymptotically flat solution
is obtained in [27], that is also derived using the [20] method. Starting
from an unusual formulation of the Lewis metric, a two parameters extension
of the method initiated in [21] allowed [28] to derive a BD version of the
Ernst equation. The method is applied to some examples, but here again, the
obtained spacetimes are generally not asymptotically flat. The matter filled
case has been considered by [29] in self interacting (ie with a potential $%
V\left( \Phi \right) $) BD, but only static axisymmetric spacetimes were
considered. Very recently, a way to generate new solutions starting from
known ones has been proposed by [30]. The technique makes use of a BD
symmetry in the traceless case ($g^{ab}T_{ab}=0$).

It is worth reminding that a scalar-metric was proposed by [31] as an SAS
vacuum BD solution. Its conformally related scalar-GR version is given in
[32]. This metric, or its [32] form, has been used in several papers to
characterize physics in a NakS (versus BH) field [32][33][34][35], and also
to suggest rotating antiscalar solutions as alternatives to Kerr BHs [36].
However, the [31] scalar-metric is actually not a vacuum BD solution, as it
is explicitly shown in [37]. The reason is that the authors of [31] applied,
without justification, to a BD spherical vacuum solution (the Brans Class I)
the Newman-Janis (NJ) algorithm, the authors (NJ) having just noticed it to
be an unexpected way to recover Kerr from Schwarzschild [38]. Let us stress
that, even in the GR framework, determining the ability of the NJ algorithm
(or some equivalent reformulations) to generate new solutions from a seed
one is not an easy task [39]. The NJ algorithm, as well as some modified
versions, nevertheless received continuous interest, not only as a way to
suggest new (generally non perfect fluid filled) solutions, but also in the
context of other theories, like supergravity. See for instance [40] for a
recent review. Related to the NJ finding [38], let us also mention that
another (but fully justified) way to derive Kerr from Schwarzschild was
obtained in [41].

In this paper, a new one to one correspondance between static axially
symmetric vacuum GR solutions and SAS massless scalar GR solutions, with
metrics having some prior form, is established. This prior form is inspired
from the metric found in [27], refered to as the SB solution in the
following. It is defined as a modification of the Kerr metric, with given $%
\left( m,a\right) $\ parameters, by inserting two unknown metric functions
in a suitable manner. Thence these functions are demanded to be such that
the scalar-Einstein field equations are satisfied. Thence, it uses Kerr as a
seed in some sense, but in a way that differs from the generating techniques
previously reviewed. The definition of some well suited $\left( \alpha
,\beta \right) $\ coordinates then allows to establish the correspondance.
Let us stress that, unlike the NJ algorithm, the method is completely
justified since it involves an explicit solving of the relevant field
equations. The SB solution is recovered as a special case, and other
explicit solutions are built, one of them being asymptotically flat. In some
sense, the introduction of the $\left( \alpha ,\beta \right) $\ coordinates
cures the problem of solving the non linear PDE system obtained in [19], for
the considered prior metric form.

\bigskip

\textit{Outline of the paper}

The prior form of the metric and the related index notations are defined in
section II. The first half of the Einstein equations is considered in II-1.
The Klein-Gordon (KG) equation is used in II-2, that suggests the definition
of the $\left( \alpha \left( r,\theta \right) ,\beta \left( r,\theta \right)
\right) $\ coordinates. In II-3, the second half of the Einstein equations
is considered. The correspondance with the general static axisymmetric
vacuum GR case is explicited in II-4. A particular asymptotically flat
solution is then considered in section III. The links with the BD theory is
discussed in section IV, while section V is dedicated to a brief conclusion.

\bigskip

\noindent \textbf{II- The considered set of axisymmetric metrics}

We consider in this paper metrics having the form 
\begin{subequations}
\begin{eqnarray}
g_{pq} &=&e^{A}k_{pq}  \label{axiform a} \\
g_{uv} &=&e^{B}k_{uv}  \label{axiform b}
\end{eqnarray}%
where $k_{ab}$ is the Kerr metric 
\end{subequations}
\begin{equation}
\left( 
\begin{array}{cccc}
k_{00} & k_{03} & 0 & 0 \\ 
k_{03} & k_{33} & 0 & 0 \\ 
0 & 0 & k_{11} & 0 \\ 
0 & 0 & 0 & k_{22}%
\end{array}%
\right) =\left( 
\begin{array}{cccc}
-V & -w\left( 1-V\right) & 0 & 0 \\ 
-w\left( 1-V\right) & 2w^{2}-w^{2}V+\Sigma \sin ^{2}\theta & 0 & 0 \\ 
0 & 0 & \frac{\Sigma }{\Delta } & 0 \\ 
0 & 0 & 0 & \Sigma%
\end{array}%
\right)  \label{Kerr metric}
\end{equation}%
in Boyer-Lindquist coordinates, and where $A$ and $B$ are $\left( r,\theta
\right) $ dependent functions. One has introduced the usual quantities%
\begin{eqnarray}
w\left( \theta \right) &=&a\sin ^{2}\theta  \label{Kerr useful} \\
\Sigma \left( r,\theta \right) &=&r^{2}+a^{2}\cos ^{2}\theta  \notag \\
\Delta \left( r\right) &=&r^{2}-2mr+a^{2}  \notag \\
V\left( r,\theta \right) &=&1-\frac{2mr}{\Sigma }.  \notag
\end{eqnarray}%
Besides the usual index convention $\left( x^{0},x^{1},x^{2},x^{3}\right)
=\left( t,r,\theta ,\phi \right) $, let us also make the convention%
\begin{eqnarray*}
\left( p,q,r,s\right) \text{ indexes } &\in &\left\{ 0,3\right\} \\
\left( u,v,w,x,y,z\right) \text{ indexes } &\in &\left\{ 1,2\right\}
\end{eqnarray*}%
while the $\left( a,b,c,d,e\right) $ indexes take the four spacetime values.
The ordering in (\ref{Kerr metric}) explicits the block diagonal structure
of Kerr's metric. The metric (\ref{axiform a}-\ref{axiform b}) is also block
diagonal, each block being "conformally" related to the Kerr corresponding
one, but with different "conformal" factors. The requirement (\ref{axiform a}%
-\ref{axiform b}) implicitly means imposing two prior relations between the
four metric functions describing the general form of an SAS metric (see for
instance eq. (1) of [42]).

The metric (\ref{axiform a}-\ref{axiform b}) is required to solve the
scalar-Einstein equation%
\begin{equation}
R_{ab}=\partial _{a}\varphi \partial _{b}\varphi  \label{scalarGR}
\end{equation}%
where $\varphi $\ is an $\left( r,\theta \right) $\ dependent scalar field.
The special case $A=B=0$ solves (\ref{scalarGR}) for $\varphi =0$, since the
Kerr metric (\ref{Kerr metric}) is Ricci flat.

The SB solution [27] corresponds to $A=0$ and $e^{B}=\left( \Delta \sin
^{2}\theta \right) ^{\sigma }$. Of course it should be recovered as a
special solution of (\ref{scalarGR}) with the corresponding scalar, that
reads $\varphi _{SB}=\sqrt{\frac{\sigma }{2}}\ln \left( \Delta \sin
^{2}\theta \right) $. This will be checked later.

From the $\left( r,\theta \right) $\ dependence of the considered
scalar-metric, one has $\partial _{p}\left( A,B,\varphi \right) =0$. The non
zero connexion components then read 
\begin{subequations}
\begin{eqnarray}
\Gamma _{qu}^{p} &=&K_{qu}^{p}+\frac{1}{2}\delta _{q}^{p}\partial _{u}A
\label{connexion a} \\
\Gamma _{pq}^{u} &=&e^{A-B}\left( K_{pq}^{u}-\frac{1}{2}k^{ux}k_{pq}\partial
_{x}A\right)   \label{connexion b} \\
\Gamma _{vw}^{u} &=&K_{vw}^{u}+\gamma _{vw}^{u}  \label{connexion c}
\end{eqnarray}%
where the $K_{ab}^{c}$\ quantities are the Kerr connexion components 
\end{subequations}
\begin{subequations}
\begin{eqnarray}
K_{qu}^{p} &=&\frac{1}{2}k^{pr}\partial _{u}k_{qr}  \label{Kerr connexion a}
\\
K_{pq}^{u} &=&-\frac{1}{2}k^{ux}\partial _{x}k_{pq}  \label{Kerr connexion b}
\\
K_{vw}^{u} &=&\frac{1}{2}k^{ux}\left( \partial _{v}k_{xw}+\partial
_{w}k_{xv}-\partial _{x}k_{vw}\right)   \label{Kerr connexion c}
\end{eqnarray}%
and where one has defined 
\end{subequations}
\begin{equation}
\gamma _{vw}^{u}=\frac{1}{2}\left( \delta _{v}^{u}\partial _{w}B+\delta
_{w}^{u}\partial _{v}B-k_{vw}k^{ux}\partial _{x}B\right) .
\label{connexion useful}
\end{equation}%
For convenience, let us introduce the following notation 
\begin{equation}
\left( c,s\right) =\left( \cos \theta ,\sin \theta \right) .
\label{cs notation}
\end{equation}

\bigskip

\noindent \textbf{II-1- The }$\left( pq\right) $\textbf{\ Einstein equation
components}

From (\ref{scalarGR}), one has $R_{pq}=0$, that writes%
\begin{equation}
\frac{1}{\sqrt{-g}}\partial _{w}\left( \sqrt{-g}\Gamma _{pq}^{w}\right)
-\Gamma _{pc}^{d}\Gamma _{qd}^{c}=0.  \label{(pq)eq1}
\end{equation}%
Using (\ref{connexion a}), (\ref{connexion b}), but also $R_{pq}\left(
k_{ab}\right) =0$ (Kerr's metric being Ricci flat) and $\partial
_{x}k_{pq}=K_{xq}^{r}k_{pr}+K_{xp}^{r}k_{qr}$\ (from the Ricci identity on
Kerr's metric), (\ref{(pq)eq1}) yields an equation that is only $A$\
dependent%
\begin{equation}
2K_{pq}^{w}\partial _{w}A-k_{pq}\left[ k^{wx}\partial _{w}A\partial _{x}A+%
\frac{1}{\sqrt{-k}}\partial _{w}\left( \sqrt{-k}k^{wx}\partial _{x}A\right) %
\right] =0.  \label{(pq)eq2}
\end{equation}%
Eliminating the bracket term thanks to the contraction by $k^{pq}$, one
obtains (since $k^{pq}k_{pq}=2$)%
\begin{equation}
\left( 2K_{pq}^{w}-k_{pq}k^{rs}K_{rs}^{w}\right) \partial _{w}A=0.
\label{(pq)eq3}
\end{equation}%
Using the Kerr metric (\ref{Kerr metric}) and the connexion components (\ref%
{connexion b}), (\ref{(pq)eq3}) yields%
\begin{equation}
k^{wx}\partial _{x}\left( \frac{k_{pq}}{\sqrt{\Delta }\left\vert
s\right\vert }\right) \partial _{w}A=0.  \label{(pq)eq4}
\end{equation}%
This equation has to be satisfied by both $k_{00}$, $k_{03}$ and $k_{33}$.
Writing out the two equations for $k_{00}$ and $k_{03}$, one obtains two
homogeneous equations on $\partial _{1}A$\ and $\partial _{2}A$. One then
shows that having a non trivial solution requires $a=0$. Thence, $A$ is
necessarily constant for $a\neq 0$. One can then set $A=0$ by redefining $B$
and the $ds^{2}$ units.

The fact that $g_{pq}=k_{pq}$ has two straigtforward consequences: (1) the
equatorial ($\theta =\pi /2$) circular orbits and their linear planar
stability (both do not involve $g_{11}$) are obtained solving the same
equations as the Kerr's case, and (2) the horizon and ergosphere are the
"same" as Kerr's%
\begin{equation}
r_{h}=m+\sqrt{m^{2}-a^{2}}  \label{horizon}
\end{equation}%
\begin{equation}
r_{e}\left( \theta \right) =m+\sqrt{m^{2}-a^{2}c^{2}}.  \label{ergosphere}
\end{equation}

To be precise, let us point out that these claims just concern the functions
that describe these orbits and surfaces in terms of $\theta $ and of the $%
\left( m,a\right) $ parameters. Indeed, the metric components $g_{uv}$\
enter their geometric and relative properties. For instance, the geometric
radial distance between two circular orbits having circumferences $C$\ and $%
C^{\prime }$\ depends on $g_{11}$.

Since $A=0$, the metric (\ref{axiform a}-\ref{axiform b}) achieves the form%
\begin{equation}
ds^{2}=k_{00}dt^{2}+2k_{03}dtd\phi +k_{33}d\phi ^{2}+e^{B}\left(
k_{11}dr^{2}+k_{22}d\theta ^{2}\right) .  \label{metric1}
\end{equation}

\bigskip

\noindent \textbf{II-2- The Klein-Gordon equation}

To pursue the integration, it would be sufficient to solve the $\left(
uv\right) $ components of (\ref{scalarGR}), since the KG equation is a
direct consequence of (\ref{scalarGR}). It is nevertheless useful to write
out the KG equation%
\begin{equation}
\partial _{a}\left( \sqrt{-g}g^{ab}\partial _{b}\varphi \right) =0.
\label{KG1}
\end{equation}%
Using (\ref{metric1}), it reads%
\begin{equation}
\partial _{1}\left( s\Delta \partial _{1}\varphi \right) +\partial
_{2}\left( s\partial _{2}\varphi \right) =0.  \label{KG2}
\end{equation}%
This form suggests defining the alternative radial coordinate%
\begin{equation}
\rho \equiv \ln \left( \frac{r-m+\sqrt{\Delta }}{\sqrt{m^{2}-a^{2}}}\right)
\label{rho def}
\end{equation}%
that yields 
\begin{subequations}
\begin{eqnarray}
\partial _{\rho } &=&\sqrt{\Delta }\partial _{1}  \label{(r versus rho)1} \\
r-m &=&\sqrt{m^{2}-a^{2}}\cosh \rho  \label{(r versus rho)2} \\
\Delta &=&\left( m^{2}-a^{2}\right) \sinh ^{2}\rho .  \label{(r versus rho)3}
\end{eqnarray}%
This allows rewriting (\ref{KG2}) in the form 
\end{subequations}
\begin{equation}
\partial _{\rho }\left( sS\partial _{\rho }\varphi \right) +\partial
_{2}\left( sS\partial _{2}\varphi \right) =0.  \label{KG3}
\end{equation}%
For convenience, one has introduced the following notation 
\begin{equation}
\left( C,S\right) =\left( \cosh \rho ,\sinh \rho \right) .
\label{CS notation}
\end{equation}%
One sees that this form returns two obvious solutions 
\begin{subequations}
\begin{eqnarray}
\varphi _{SB} &=&\Lambda \ln \left( Ss\right)  \label{KG sol 1} \\
\varphi _{2} &=&\Lambda Cc  \label{KG sol 2}
\end{eqnarray}%
where $\Lambda $\ is any constant, and $\varphi _{SB}$\ the scalar entering
the SB solution [27]. The form of these two solutions suggests defining new $%
\left( \alpha ,\beta \right) $ coordinates by 
\end{subequations}
\begin{equation}
\left( \alpha ,\beta \right) =\left( Ss,Cc\right) .  \label{alpha-beta def}
\end{equation}%
In terms of the initial Boyer-Lindquist coordinates, these coordinates read 
\begin{subequations}
\begin{eqnarray}
\alpha &=&\frac{\sqrt{r^{2}-2mr+a^{2}}}{\sqrt{m^{2}-a^{2}}}\sin \theta
\label{alpha(r,theta)} \\
\beta &=&\frac{r-m}{\sqrt{m^{2}-a^{2}}}\cos \theta  \label{beta(r,theta)}
\end{eqnarray}%
and behave as $\left( r\sin \theta ,r\cos \theta \right) $ for $%
r\longrightarrow \infty $, up to the $\left( m^{2}-a^{2}\right) ^{-1/2}$
factor. From (\ref{alpha-beta def}) and the trigonometric identities $%
c^{2}+s^{2}=1$\ and $C^{2}-S^{2}=1$, one obtains the following relations 
\end{subequations}
\begin{eqnarray}
\left[ \left( Cs\right) ^{2}+\left( Sc\right) ^{2}\right] \partial _{\alpha
}\left( C,S\right) &=&Cs\left( S,C\right)  \label{trigo deriv} \\
\left[ \left( Cs\right) ^{2}+\left( Sc\right) ^{2}\right] \partial _{\alpha
}\left( c,s\right) &=&Sc\left( -s,c\right)  \notag \\
\left[ \left( Cs\right) ^{2}+\left( Sc\right) ^{2}\right] \partial _{\beta
}\left( C,S\right) &=&Sc\left( S,C\right)  \notag \\
\left[ \left( Cs\right) ^{2}+\left( Sc\right) ^{2}\right] \partial _{\beta
}\left( c,s\right) &=&Cs\left( s,-c\right)  \notag
\end{eqnarray}%
that turn out to be useful in the calculations to do. The derivation
operators transform as%
\begin{eqnarray}
\partial _{\rho } &=&Cs\partial _{\alpha }+Sc\partial _{\beta }
\label{deriv transfo} \\
\partial _{2} &=&Sc\partial _{\alpha }-Cs\partial _{\beta }.  \notag
\end{eqnarray}%
Reinserting in (\ref{KG3}) returns, after a lengthy calculation, and using (%
\ref{trigo deriv}), the KG equation in the nice form%
\begin{equation}
\partial _{\alpha }\left( \alpha \partial _{\alpha }\varphi \right) +\alpha
\partial _{\beta }\partial _{\beta }\varphi =0.  \label{KG4}
\end{equation}%
This form points out two other obvious solutions, besides (\ref{KG sol 1})
and (\ref{KG sol 2}) 
\begin{subequations}
\begin{eqnarray}
\varphi _{3} &=&\Lambda \beta \ln \alpha  \label{KG sol 3} \\
\varphi _{N} &=&\frac{\Lambda }{\sqrt{\alpha ^{2}+\beta ^{2}}}.
\label{KG sol 4}
\end{eqnarray}%
The solution (\ref{KG sol 3}) is nothing but the product of (\ref{KG sol 1})
and (\ref{KG sol 2}), that turns out to be a solution too. The solution (\ref%
{KG sol 4}) results from the fact that (\ref{KG4}) is the classical
Laplacian equation written in cylindrical coordinates, in the case of an
axisymmetric potential. For this reason, we will refer to (\ref{KG sol 4})
as being the Newtonian scalar.

It may be worth spoting that if $\varphi $\ solves (\ref{KG4}), so do its
successive derivatives with respect to $\beta $. (Remark that $\varphi
_{SB}=\partial _{\beta }\varphi _{3}$.) Along these lines, new solutions can
also be obtained by integration with respect to $\beta $. For instance 
\end{subequations}
\begin{subequations}
\begin{eqnarray}
\varphi _{5} &=&\Lambda \ln \left( \beta +\sqrt{\alpha ^{2}+\beta ^{2}}%
\right)  \label{KG sol 5} \\
\varphi _{6} &=&\Lambda \left[ \beta \ln \left( \beta +\sqrt{\alpha
^{2}+\beta ^{2}}\right) -\sqrt{\alpha ^{2}+\beta ^{2}}\right]
\label{KG sol 6}
\end{eqnarray}%
also solve (\ref{KG4}), and are related to $\varphi _{N}$\ by $\varphi
_{N}=\partial _{\beta }\varphi _{5}=\partial _{\beta }\partial _{\beta
}\varphi _{6}$.

See the appendix for a quicker demonstration of (\ref{KG4}), using the $%
\left( \alpha ,\beta \right) $\ coordinates.

\bigskip

\noindent \textbf{II-3- The }$\left( uv\right) $\textbf{\ Einstein equation
components}

From (\ref{scalarGR}), one has $R_{uv}=\partial _{u}\varphi \partial
_{v}\varphi $, that writes 
\end{subequations}
\begin{equation}
\frac{1}{\sqrt{-g}}\partial _{w}\left( \sqrt{-g}\Gamma _{uv}^{w}\right)
-\partial _{u}\partial _{v}\ln \sqrt{-g}-\Gamma _{up}^{q}\Gamma
_{vq}^{p}-\Gamma _{ux}^{y}\Gamma _{vy}^{x}=\partial _{u}\varphi \partial
_{v}\varphi .  \label{(uv)eq1}
\end{equation}%
Using (\ref{connexion a}), (\ref{connexion c}) and (\ref{connexion useful}),
with $A=0$, but also $R_{uv}\left( k_{ab}\right) =0$ (Kerr's metric is Ricci
flat) and $\partial _{y}k_{uv}=K_{yu}^{x}k_{vx}+K_{yv}^{x}k_{ux}$\ (from the
Ricci identity on Kerr's metric), (\ref{(uv)eq1}) yields%
\begin{equation}
K_{vp}^{p}\partial _{u}B+K_{up}^{p}\partial _{v}B-\frac{1}{\sqrt{-k}}%
k_{uv}\partial _{w}\left( \sqrt{-k}k^{wz}\partial _{z}B\right) =2\partial
_{u}\varphi \partial _{v}\varphi .  \label{(uv)eq2}
\end{equation}%
Expliciting the $\left( 11\right) $, $\left( 12\right) $ and $\left(
22\right) $\ components yields, using the $\rho $ radial coordinate and (\ref%
{(r versus rho)1}) 
\begin{subequations}
\begin{eqnarray}
Cs\partial _{\rho }B-Ss\partial _{\rho }\partial _{\rho }B-S\partial
_{2}\left( s\partial _{2}B\right) &=&2Ss\left( \partial _{\rho }\varphi
\right) ^{2}  \label{(uv)eq3 1} \\
Sc\partial _{\rho }B+Cs\partial _{2}B &=&2Ss\partial _{\rho }\varphi
\partial _{2}\varphi  \label{(uv)eq3 2} \\
2Sc\partial _{2}B-S\partial _{2}\left( s\partial _{2}B\right) -Ss\partial
_{\rho }\partial _{\rho }B-Cs\partial _{\rho }B &=&2Ss\left( \partial
_{2}\varphi \right) ^{2}.  \label{(uv)eq3 3}
\end{eqnarray}%
Let us now rewrite these equations in $\left( \alpha ,\beta \right) $\
coordinates. Combining the equation (\ref{(uv)eq3 2}), and the difference of
(\ref{(uv)eq3 1}) and (\ref{(uv)eq3 3}), returns 
\end{subequations}
\begin{subequations}
\begin{eqnarray}
\partial _{\alpha }B &=&\alpha \left[ \left( \partial _{\alpha }\varphi
\right) ^{2}-\left( \partial _{\beta }\varphi \right) ^{2}\right]
\label{(uv)eq4 1} \\
\partial _{\beta }B &=&2\alpha \partial _{\alpha }\varphi \partial _{\beta
}\varphi .  \label{(uv)eq4 2}
\end{eqnarray}%
It appears that the integrability condition of (\ref{(uv)eq4 1}) and (\ref%
{(uv)eq4 2}) is ensured by the KG equation (\ref{KG4}). An $\left( uv\right) 
$\ equation remains to be written, that can be built from the sum of (\ref%
{(uv)eq3 1}) and (\ref{(uv)eq3 3}). It turns out that this equation is
solved thanks to (\ref{KG4}).

See the appendix for a quicker demonstration of these results, using the $%
\left( \alpha ,\beta \right) $\ coordinates from the start.

\bigskip

\noindent \textbf{II-4- Generating solutions}

From the previous sections, a $\left( \varphi ,B\right) $\ solution can be
obtained by (1) solving first the KG equation (\ref{KG4}), and then (2)
integrating the system (\ref{(uv)eq4 1})-(\ref{(uv)eq4 2}). There is then a
direct correspondance with the issue of seeking a static axisymmetric
solution of vacuum GR. Indeed, such a metric can be written [2] 
\end{subequations}
\begin{equation}
ds^{2}=-e^{2U}dt^{2}+e^{-2U}\left[ e^{2\gamma }\left( d\rho
^{2}+dz^{2}\right) +\rho ^{2}d\phi ^{2}\right]  \label{vac axial static}
\end{equation}%
where the metric functions $U\left( \rho ,z\right) $ and $\gamma \left( \rho
,z\right) $ have to solve 
\begin{subequations}
\begin{eqnarray}
\partial _{\rho }\left( \rho \partial _{\rho }U\right) +\rho \partial
_{z}\partial _{z}U &=&0  \label{vac axial static eq 1} \\
\partial _{\rho }\gamma &=&\rho \left[ \left( \partial _{\rho }U\right)
^{2}-\left( \partial _{z}U\right) ^{2}\right]  \label{vac axial static eq 2}
\\
\partial _{z}\gamma &=&2\rho \partial _{\rho }U\partial _{z}U.
\label{vac axial static eq 3}
\end{eqnarray}%
Thence, any $\left( U\left( \rho ,z\right) ,\gamma \left( \rho ,z\right)
\right) $ solution of (\ref{vac axial static eq 1})-(\ref{vac axial static
eq 3}) immediately results in a $\left( \varphi \left( \alpha ,\beta \right)
,B\left( \alpha ,\beta \right) \right) $\ solution of the problem considered
in this paper, by just making the changes $\left( \rho ,z\right) \rightarrow
\left( \alpha ,\beta \right) $ and $\left( U,\gamma \right) \rightarrow
\left( \varphi ,B\right) $. There is then a one to one correspondance
between these two problems. Let us stress that the $\left( \alpha ,\beta
\right) $ definition depends on the Kerr's mass and angular momentum, in
such a way that the correspondance works for any $\left( m,a\right) $\
parameters. Note that while both problems are GR issues, the former
corresponds to a static field in a vacuum spacetime, while the later to a
rotating field in a not vacuum (but massless scalar filled) spacetime.

Considering the four first solutions of the KG equation obtained in II-2,
one obtains 
\end{subequations}
\begin{subequations}
\begin{eqnarray}
\varphi _{SB} &=&\Lambda \ln \alpha \text{ \ \ }\Longrightarrow \text{ \ \ }%
B_{SB}=\Lambda ^{2}\ln \alpha  \label{sol 1} \\
\varphi _{2} &=&\Lambda \beta \text{ \ \ }\Longrightarrow \text{ \ \ }%
B_{2}=-\Lambda ^{2}\frac{\alpha ^{2}}{2}  \label{sol 2} \\
\varphi _{3} &=&\Lambda \beta \ln \alpha \text{ \ \ }\Longrightarrow \text{
\ \ }B_{3}=\Lambda ^{2}\left( \beta ^{2}\ln \alpha -\frac{1}{4}\alpha ^{2}%
\left[ 1-2\ln \alpha +2\left( \ln \alpha \right) ^{2}\right] \right)
\label{sol 3} \\
\varphi _{N} &=&\frac{\Lambda }{\sqrt{\alpha ^{2}+\beta ^{2}}}\text{ \ \ }%
\Longrightarrow \text{ \ \ }B_{N}=-\frac{\Lambda ^{2}\alpha ^{2}}{2\left(
\alpha ^{2}+\beta ^{2}\right) ^{2}}  \label{sol 4}
\end{eqnarray}%
The $\left( \varphi _{SB},B_{SB}\right) $ solution is the SB solution [27].
It is clear from (\ref{alpha(r,theta)})-(\ref{beta(r,theta)}) and (\ref%
{metric1}) that the SB, the $\left( \varphi _{2},B_{2}\right) $ and the $%
\left( \varphi _{3},B_{3}\right) $ solutions are not asymptotically flat, a
point that makes their astrophysical usefulness debatable. On the other
hand, the Newtonian solution $\left( \varphi _{N},B_{N}\right) $ is
asymptotically flat.

From the previously pointed out correspondance, the SB solution is
associated to the Minkowski spacetime (in some non cartesian coordinates),
while the Newtonian one is associated to the Curzon-Chazy spacetime [2].

It is worth also having a look on the fifth solution of the KG equation
obtained in II-2. Integrating for $B_{5}$ yields 
\end{subequations}
\begin{equation}
B_{5}=2\Lambda ^{2}\ln \left( 1+\frac{\beta }{\sqrt{\alpha ^{2}+\beta ^{2}}}%
\right)  \label{sol 5}
\end{equation}%
ie, in terms of $\left( r,\theta \right) $ coordinates, using (\ref%
{alpha-beta def}) 
\begin{equation}
e^{B_{5}}=\left( 1+\frac{\left( r-m\right) c}{\sqrt{r^{2}-2mr+a^{2}+\left(
m^{2}-a^{2}\right) c^{2}}}\right) ^{2\Lambda ^{2}}  \label{sol 5bis}
\end{equation}%
This, with (\ref{metric1}), suggests that the spacetime is not
asymptotically flat. From the previous correspondance, the $\left( \varphi
_{5},B_{5}\right) $ solution is associated to the Gautreau-Hoffman spacetime
[2].

The $\left( \varphi _{6},B_{6}\right) $ solution is obviously not
asymptotically flat. Indeed, (\ref{scalarGR}) shows that it is even not
Ricci flat at infinity, since $\partial _{\alpha }\varphi _{6}=-\Lambda
\alpha \left( \beta +\sqrt{\alpha ^{2}+\beta ^{2}}\right) ^{-1}$ and $%
\partial _{\beta }\varphi _{6}=\varphi _{5}=\Lambda \ln \left( \beta +\sqrt{%
\alpha ^{2}+\beta ^{2}}\right) $ do not vanish at infinity.

\bigskip

\noindent \textbf{III- The Newtonian solution}

Since it is asymptotically flat, let us have a closer look on the Newtonian
solution.\ Its metric (\ref{sol 4}) explicitly reads 
\begin{equation}
ds_{N}^{2}=k_{00}dt^{2}+2k_{03}dtd\phi +k_{33}d\phi ^{2}+\exp \left( -\frac{%
\Lambda ^{2}\left( m^{2}-a^{2}\right) \left( r^{2}-2mr+a^{2}\right) s^{2}}{2%
\left[ r^{2}-2mr+a^{2}+\left( m^{2}-a^{2}\right) c^{2}\right] ^{2}}\right)
\left( k_{11}dr^{2}+k_{22}d\theta ^{2}\right) .  \label{Kerr-scalar}
\end{equation}%
The $g_{00}$ and $g_{03}$ metric tensor components being identical to Kerr,
the $m$ and $a$ constants have the same ADM mass and angular momentum
meanings. As already mentionned, the Kerr horizon (\ref{horizon}) and
ergosphere (\ref{ergosphere}) surfaces are recovered, since they just depend
on the $g_{pq}$ metric components, which are the same as Kerr.

The scalar associated to (\ref{Kerr-scalar}) reads, in terms of $\left(
r,\theta \right) $ coordinates%
\begin{equation}
\varphi _{N}=\frac{\Lambda \sqrt{m^{2}-a^{2}}}{\sqrt{r^{2}-2mr+a^{2}+\left(
m^{2}-a^{2}\right) c^{2}}}.  \label{Kerr-scalar scalar}
\end{equation}%
The explicit $\theta $ dependence of $\varphi _{N}$ shows that the (\ref%
{Kerr-scalar})-(\ref{Kerr-scalar scalar}) solution is not the BD-Kerr
solution obtained in [19] (equations (24)-(26) of [19]).

\bigskip

\noindent \textbf{III-1- Singularities}

One knows that the Kerr metric is singularity free outside its (external)
horizon (\ref{horizon}), and that the horizon itself is a regular surface.
However, from the Hawking theorem [16], this should not be true for the
spacetime (\ref{Kerr-scalar}) since $\partial \varphi _{N}\neq 0$ while (1)
it is asymptotically flat and, (2) its horizon surface is finite. Indeed,
the scalar curvature reads, from (\ref{scalarGR})%
\begin{equation}
R=g^{uv}\partial _{u}\varphi _{N}\partial _{v}\varphi _{N}
\label{scalar curv N 1}
\end{equation}%
which, from (\ref{sol 4}), yields%
\begin{equation}
R=\frac{16\Lambda ^{2}}{\Sigma }\frac{C^{2}S^{2}+c^{2}s^{2}}{\left(
S^{2}s^{2}+C^{2}c^{2}\right) ^{2}}\exp \left( \frac{\Lambda ^{2}S^{2}s^{2}}{%
2\left( S^{2}s^{2}+C^{2}c^{2}\right) ^{2}}\right) .  \label{scalar curv N 2}
\end{equation}%
One sees that (\ref{scalar curv N 2}) diverges whatever the way $%
s^{2}S^{2}+c^{2}C^{2}\longrightarrow 0$, ie the way $\left( c,S\right)
\longrightarrow \left( 0,0\right) $ (and only in this case for $r\geq r_{h}$%
). The points having $\theta =\pi /2$ and $\rho =0$ are then NakS points.\
Thence, the horizon is not regular everywhere, since its equator $\theta
=\pi /2$, and the equator only (considering points having $r\geq r_{h}$), is
scalar curvature singular.

However, the fact that the scalar curvature does not diverge on the points
having $\left( r\geq r_{h},\theta \right) \neq \left( r_{h},\pi /2\right) $
is not sufficient to prove that none of these points is singular. A quantity
that is often regarded as a reliable singularity indicator is the
Kretschmann scalar%
\begin{equation}
\widehat{K}\equiv R_{abcd}R^{abcd}  \label{Kretschmann 1}
\end{equation}%
where $R_{abcd}$\ is the Riemann-Christoffel curvature tensor. From (\ref%
{connexion a})-(\ref{connexion c}), one finds that%
\begin{eqnarray}
R_{pqrs} &=&e^{-B}Q_{pqrs}  \label{RC} \\
R_{pquv} &=&Q_{pquv}  \notag \\
R_{puqv} &=&Q_{puqv}+q_{puqv}  \notag \\
R_{uvwx} &=&e^{B}\left( Q_{uvwx}+q_{uvwx}\right)  \notag \\
R_{pqru} &=&R_{puvw}=0  \notag
\end{eqnarray}%
where $Q_{abcd}$\ is the Kerr's metric Riemann-Christoffel curvature tensor,
and%
\begin{eqnarray}
q_{puqv} &=&k_{pr}K_{qw}^{r}\gamma _{uv}^{w}  \label{RC useful} \\
q_{uvwx} &=&k_{uy}\left( \partial _{w}\gamma _{vx}^{y}-\partial _{x}\gamma
_{vw}^{y}+K_{vx}^{z}\gamma _{wz}^{y}+K_{wz}^{y}\gamma
_{vx}^{z}-K_{vw}^{z}\gamma _{xz}^{y}-K_{xz}^{y}\gamma _{vw}^{z}+\gamma
_{wz}^{y}\gamma _{vx}^{z}-\gamma _{xz}^{y}\gamma _{vw}^{z}\right) .  \notag
\end{eqnarray}%
If $B=0$, which implies $\gamma _{vw}^{u}=0$ from (\ref{connexion useful}), $%
\widehat{K}$ is finite since the Kerr's metric is regular in the considered
region. Any divergence of $\widehat{K}$ can then only appear from the $%
\gamma _{vw}^{z}$ \ and $\partial \gamma _{vw}^{z}$, ie from the $\partial
B_{N}$\ and $\partial \partial B_{N}$, quantities entering the $q_{abcd}$\
terms in (\ref{RC}). It is then easy to see from (\ref{sol 4}) that a
divergence of $\widehat{K}$ can only occur at points where $c=S=0$. The set
of Kretschmann singularity points is then included in the scalar curvature
points set. \ This strongly suggests that the horizon is regular at points
not belonging to the equatorial NakS ring $\left( r,\theta \right) =\left(
r_{h},\pi /2\right) $\footnote{%
Extending to the $r<r_{h}$ region, one sees on (\ref{scalar curv N 2}) that $%
\Sigma =0$ is a scalar curvature singularity of the considered spacetime (it
is a singularity, but not a scalar curvature one, in Kerr's spacetime). The
spacetime has then two ring singularities: one is naked, the second being
hidden (the latter being the counterpart of the usual Kerr's, in some sense).%
}.

Let us stress that it was recently found by [43] that the (not
asymptotically flat) SB solution, ie (\ref{sol 1}), also exhibits such a
ringlike NakS structure.

\bigskip

\noindent \textbf{III-2- The }$a=0$\textbf{\ subcase (static case)}

Let us now specify the Newtonian solution (\ref{sol 4}) to the $a=0$ case.
The vacuum Kerr's solution returns the (vacuum) spherical Schwarzschild
solution for $a=0$. In the non vacuum, but massless scalar filled, case, a
spherical solution is known, often named the JNW metric, refering to the
Janis-Newman-Winicour 1968's paper [44]. It is worth to point out that this
solution was in fact earlier discovered by Fisher in his 1948 paper [45],
but using an areal radial coordinate. It seems then fair to use the FJNW
acronym when refering to this solution, and so will I do in this paper. The
FJNW solution includes Schwarzschild as a limit (non scalarized) case. It is
then naturel to suspect the existence of a massless scalar filled GR
solution, that would depend on a "rotation parameter" $a$, and that would
return (1) the Kerr spacetime for a vanishing scalar, and (2) the FJNW
solution for $a=0$.

The solution (\ref{sol 4}) indeed returns the Kerr spacetime for a vanishing
scalar, ie for $\Lambda =0$. On the other hand, making $a=0$\ in (\ref%
{Kerr-scalar}) does not return the FJNW solution, but%
\begin{equation}
ds^{2}=-\left( 1-\frac{2m}{r}\right) dt^{2}+\exp \left( -\frac{\Lambda
^{2}m^{2}\left( r^{2}-2mr\right) s^{2}}{2\left( r^{2}-2mr+m^{2}c^{2}\right)
^{2}}\right) \left[ \left( 1-\frac{2m}{r}\right) ^{-1}dr^{2}+r^{2}d\theta
^{2}\right] +r^{2}s^{2}d\phi ^{2}.  \label{Kerr-scalar a=0}
\end{equation}%
This static but non spherical solution is known from long, see equations
(17)-(18) of Penney's paper [46]. From (\ref{Kerr-scalar scalar}), the
scalar field reads%
\begin{equation}
\varphi =\frac{m\Lambda }{\sqrt{r^{2}-2mr+m^{2}c^{2}}}.
\label{Kerr-scalar a=0 scalar}
\end{equation}%
The metric (\ref{Kerr-scalar a=0}) has an horizon, that reads%
\begin{equation}
r=2m  \label{horizon a=0}
\end{equation}%
and whose equator's points $\theta =\pi /2$, and only these points, are
singular, as it can be directly checked from the $\left(
S^{2}s^{2}+C^{2}c^{2}\right) $ expression entering (\ref{scalar curv N 2}),
that reads for $a=0$%
\begin{equation}
S^{2}s^{2}+C^{2}c^{2}=\left( \frac{r}{m}-1\right) ^{2}-s^{2}.
\label{ring a=0}
\end{equation}%
This agrees with the fact that the metric, while static since $g_{03}=0$, is
not spherical because of the presence of the ($\theta $ dependent)
exponential in front of the bracket in (\ref{Kerr-scalar a=0}).
Incidentally, this also shows that (\ref{Kerr-scalar a=0}) can not be FJNW
in deguise. The ring character of the NakS, stuck on the horizon, is
obviously a property inherited from the general $a\neq 0$\ case.

The fact that the FJNW metric does not appear as a subcase of (\ref%
{Kerr-scalar a=0}) means that (\ref{Kerr-scalar}) can not be interpreted as
a "rotating version of FJNW". It also\ strongly suggests the existence of
asymptotically flat Kerr like solutions of (\ref{scalarGR}), that do not
fulfil (\ref{axiform a}). Indeed, recovering FJNW as a rotationless limit
case, with its properties ([15]), is incompatible with the $A=0$ conclusion
obtained in II-1.

\bigskip

\noindent \textbf{III-3- Orbits in the equatorial plane of the static
solution}

As already mentioned, the presence of the scalar field affects neither the
existence condition of equatorial circular orbits nor their linear stability
(using the radial coordinate defined by the form of the metric (\ref{metric1}%
)). The situation is very similar to the case of a scalarized version of the 
$\gamma $-metric reported in [47]. However, this does not mean that the
physics is unaffected by the scalar, even in the equatorial plane. The aim
of this subsection is to illustrate this point. For convenience, we specify
to the static case, that makes all the calculations easily tractable.

In the $\theta =\pi /2$ plane, the metric (\ref{Kerr-scalar a=0}) simplifies
into%
\begin{equation}
ds^{2}=-\left( 1-\frac{2m}{r}\right) dt^{2}+\exp \left( -\frac{\Lambda
^{2}m^{2}}{2\left( r^{2}-2mr\right) }\right) \left( 1-\frac{2m}{r}\right)
^{-1}dr^{2}+r^{2}d\phi ^{2}.  \label{Kerr-scalar a=0 equat}
\end{equation}%
One has already pointed out that the exponential term in $g_{11}$ impacts
the distance between close circular orbits. The effect is mostly important
near the $r=2m$ NakS, where the distance between two close orbits of radii $%
r $ and $r+dr$\ goes to zero, because of the vanishing of the exponential
factor. Related to this, the (coordinate) time needed for a photon to
radially propagate from $r=2m$ to an external observer at coordinate $%
r_{obs} $, that reads%
\begin{equation}
\Delta t\left( r=2m\rightarrow r_{obs}\right) =\int_{2m}^{r_{obs}}\frac{rdr}{%
r-2m}\exp \left( -\frac{\Lambda ^{2}m^{2}}{4\left( r^{2}-2mr\right) }\right)
,  \label{time flying convergence}
\end{equation}%
converges for $\Lambda \neq 0$, unlike what happens in the Schwarzschild
case. On the other hand, the convergence does not occur for a radial polar
photon ($\theta =0$ or $\pi $), since the argument of the exponential term
in (\ref{Kerr-scalar a=0}) cancels then. More generally, the convergence
does not occur for any photon crossing the Penney's $r=2m$ sphere at any non
equatorial point, since then $c\neq 0$, so that the exponential goes to $1$
when $r\longrightarrow 2m$. However, let us also spot that despite the
convergence of (\ref{time flying convergence}), the far observed behaviour
of a clock at rest close to the Penney's sphere is frozen, even in the
equatorial plane, since $g_{00}\left( r=2m\right) =0$ for any $\theta $, the 
$\pi /2$ case included. The consequence is that in the equatorial plane,
while circular close to the NakS (non geodesic) motions are frozen (the
areal NakS circumference being $4\pi m$), radial infallings are not.

Let us point out that the same observations, ie (1) (coordinate) time
convergence for a radial photon propagating from any NakS point, but (2)
frozen rest clock behaviour near any NakS point, also occurs in the FJNW
spherical metric [44][45].\ This is obvious from the isotropic form of this
metric [15]%
\begin{equation}
ds^{2}=-\left( \frac{r-k}{r+k}\right) ^{2\lambda }dt^{2}+\left( \frac{r+k}{r}%
\right) ^{4}\left( \frac{r-k}{r+k}\right) ^{2-2\lambda }\left(
dr^{2}+r^{2}d\Omega ^{2}\right)  \label{FJNW equatorial}
\end{equation}%
where $\lambda \in \left] 0,1\right[ $ ($\lambda =1$ corresponding to the GR
Schwarzschild metric)\footnote{%
Let us point out that in the FJNW case, the fact that close to the NakS rest
clocks are frozen does not mean that circular motions are frozen, since the
NakS circumference is zero. Indeed, close to NakS circular geodesics (that
exist for $\lambda <1/2$) return a divergent far observed frequency [15].}.
At least for these two metrics, the time convergence for a photon reaching
the horizon only concerns orbits approaching points belonging to the NakS
location.

\bigskip

\noindent \textbf{IV- Brans-Dicke vacuum solutions}

It is well-known [48] that in a four-dimensional spacetime, the conformal
transformation%
\begin{equation}
g_{ab}=\Phi \overline{g}_{ab}  \label{conf transfo}
\end{equation}%
yields, for any constant $\omega $ (supposed to be $>-3/2$) 
\begin{equation}
\int \left[ \Phi \overline{R}-\frac{\omega }{\Phi }\left( \overline{\partial 
}\Phi \right) ^{2}\right] \sqrt{-\overline{g}}d^{4}x=\int \left( R-\frac{1}{2%
}\left( \partial \varphi \right) ^{2}\right) \sqrt{-g}d^{4}x
\label{conf related theories}
\end{equation}%
where $\Phi $ is any positive scalar function and 
\begin{equation}
\varphi =\sqrt{2\omega +3}\ln \Phi .  \label{phi def}
\end{equation}%
This means that the vacuum BD action of the BD gravitational field $\left(
\Phi ,\overline{g}_{ab}\right) $ identifies with the GR action, with
gravitational field $g_{ab}$, but filled by the (matter source) massless
scalar $\varphi $. Thence, any solution of (\ref{scalarGR}) is conformally
associated to a vacuum BD solution [18]. This ($\omega $)BD solution reads 
\begin{subequations}
\begin{eqnarray}
\Phi &=&\exp \left( \frac{\varphi }{\sqrt{2\omega +3}}\right)
\label{BD sol scalar} \\
\overline{g}_{ab} &=&\exp \left( -\frac{\varphi }{\sqrt{2\omega +3}}\right)
g_{ab}.  \label{BD sol metric}
\end{eqnarray}%
It is then possible to built a vacuum SAS BD solution from any vacuum static
axisymmetric GR solution, using first the correspondance reported in II-4.
Let us mention that the Kerrlike BD solution built this way from the
Newtonian solution (\ref{sol 4}), or (\ref{Kerr-scalar})-(\ref{Kerr-scalar
scalar}), is also asymptotically flat, since $\varphi _{N}$\ vanishes in far
regions.

Experiments strongly constrain BD/ST theories to satisfy $\omega
_{0}>4.10^{5}$, where $\omega _{0}$ is the BD parameter, or the present
value of $\omega \left( \Phi \right) $ in the ST case [8]. In such
circumstances, vacuum BD gravity, and also to some extent ST gravity, is
asymptotically equivalent to massless scalar filled GR [15][28][49]. In
other words, the solutions obtained solving (\ref{KG4}) and (\ref{(uv)eq4 1}%
)-(\ref{(uv)eq4 2}) can directly serve as vacuum BD/ST solutions in the
large $\omega $ case, without having to explicitly consider the conformal
correspondance (\ref{BD sol scalar})-(\ref{BD sol metric}). Linked to this,
let us remind that for any scalar function $\varphi $ chosen "independently
on $\omega $", (\ref{BD sol scalar}) yields 
\end{subequations}
\begin{equation}
\Phi =1+\frac{\varphi }{\sqrt{2\omega }}+O\left( \frac{1}{\omega }\right) .
\label{large omega scalar}
\end{equation}%
Despite that $\Phi $ goes to a constant value, the $\frac{\omega }{\Phi ^{2}}%
\partial _{a}\Phi \partial _{b}\Phi $ term entering the full BD equation
does then not vanish in the large $\omega $\ limit, but results in a $%
\partial _{a}\varphi \partial _{b}\varphi $ contribution. This is coherent
with the fact that a vacuum BD solution does not reduce to a GR vacuum
solution (that would have been Kerr in the SAS case) in the "$\omega
\longrightarrow \infty $ limit" [49].

Let us remark that while the massive geodesics are not the same in the
scalar-Einstein and in the corresponding BD solutions, they are
asymptotically identical in the $\omega \longrightarrow \infty $ limit,
since the conformal factor is constant in this limit.

\bigskip

\noindent \textbf{V- Conclusion and outlook}

One has established a non trivial one to one correspondance that allows to
built an SAS massless scalar-GR solution from any static axisymmetric vacuum
GR solution. While this in turn also results in a way to obtain an SAS
vacuum BD solution, one has pointed out that the so obtained massless
scalar-GR solutions can also directly serve as a (large $\omega $) vacuum
BD/ST solutions for practical purposes.

It was claimed in [25] that a Kerr-like BD solution should: (1) only depend
on the three $\left( m,a,\omega \right) $ parameters, (2) go to Kerr
spacetime in the $\omega \longrightarrow \infty $\ limit, and (3) return
Schwarzschild's spacetime for $a=0$. From the results obtained in this
paper, it seems that the truth is by far more complex. The solutions
explicitly presented in this paper can be interpreted as limit of BD
solutions for $\omega \longrightarrow \infty $, but they differ from Kerr
spacetime. The particular case (\ref{Kerr-scalar a=0}) of the Newtonian
solution has $a=0$\ but differs from Schwarzschild. The fact that the
Newtonian solution (\ref{Kerr-scalar}), for instance, depends on a parameter 
$\Lambda $, besides $\left( m,a\right) $, shows that the ($\omega $)BD
solution built from it using (\ref{BD sol scalar})-(\ref{BD sol metric})
results in a family of Kerr-like solutions that depends on more than the
three $\left( m,a,\omega \right) $\ parameters.

Let us mention that a lot of authors explicitly or implicitly still suppose
that (2) is true (see for instance [26] or [27]). Such a presupposition
results in a drastic impoverishment of the (large $\omega $) ST potential
predictions. Among these the drastically different behaviour of far observed
orbital frequencies in the spherical case with respect to GR [15]. It would
be of interest to know whether such a different from GR behaviour could also
occur in the rotating case. This requires exploring more general SAS
solutions of (\ref{scalarGR}) with $g_{pq}\neq k_{pq}$, which means ruling
out the restrictive hypothesis (\ref{axiform a}).

\bigskip

\noindent \textbf{Acknowledgments}

I would like to thank here an anonymous referee for several very interesting
historical references, especially concerning the often named JNW (but
accordingly here renamed FJNW) scalar-Einstein solution.

\bigskip

\noindent \textbf{Appendix: calculations using the }$\left( \alpha ,\beta
\right) $\textbf{\ coordinates}

Using (\ref{rho def}) and (\ref{alpha-beta def}), one obtains, in the $%
\left( \widetilde{x}^{a}\right) \equiv \left( t,\alpha ,\beta ,\phi \right) $
coordinates%
\begin{equation}
ds^{2}=\widetilde{g}_{ab}d\widetilde{x}^{a}d\widetilde{x}%
^{b}=k_{00}dt^{2}+2k_{03}dtd\phi +k_{33}d\phi ^{2}+G\left( d\alpha
^{2}+d\beta ^{2}\right)  \label{metric2}
\end{equation}%
where%
\begin{equation}
G=\frac{\Sigma e^{B}}{S^{2}c^{2}+C^{2}s^{2}}.  \label{G def}
\end{equation}%
From (\ref{alpha-beta def}) and the trigonometric identities, one could
explicitly obtain $\left( c,s,C,S\right) $ in terms of $\left( \alpha ,\beta
\right) $, but this is not needed here. Indeed, since $\sqrt{-\widetilde{g}}%
=Gs\sqrt{\Delta }=\sqrt{m^{2}-a^{2}}\alpha G$, the Dalembertian operator
reads%
\begin{eqnarray}
\square \varphi &=&\frac{1}{\sqrt{-\widetilde{g}}}\left[ \partial _{\alpha
}\left( \sqrt{-\widetilde{g}}g^{\alpha \alpha }\partial _{\alpha }\varphi
\right) +\partial _{\beta }\left( \sqrt{-\widetilde{g}}g^{\beta \beta
}\partial _{\beta }\varphi \right) \right]  \label{Dalemb(alpha-beta)} \\
&=&\frac{\sqrt{m^{2}-a^{2}}}{\sqrt{-\widetilde{g}}}\left[ \partial _{\alpha
}\left( \alpha \partial _{\alpha }\varphi \right) +\alpha \partial _{\beta
}\partial _{\beta }\varphi \right]  \notag
\end{eqnarray}%
from which one directly obtains (\ref{KG4}).

Writing now (\ref{metric2}) in the form%
\begin{equation}
ds^{2}=k_{00}dt^{2}+2k_{03}dtd\phi +k_{33}d\phi ^{2}+He^{B}\left( d\alpha
^{2}+d\beta ^{2}\right)  \label{metric3}
\end{equation}%
(the $B=0$ version of which being Kerr in $\left( \widetilde{x}^{a}\right) $%
\ coordinates), the $\widetilde{\Gamma }_{\ast \ast }^{\ast }$ connexion
components achieve a form like (\ref{connexion a})-(\ref{connexion c}) with $%
A=0$, but with $\left( \alpha ,\beta \right) $ instead of $\left( r,\theta
\right) $. Thence, (\ref{(uv)eq2}) is replaced by%
\begin{equation}
\widetilde{K}_{vp}^{p}\partial _{u}B+\widetilde{K}_{up}^{p}\partial _{v}B-%
\frac{1}{\alpha He^{B}}k_{uv}\left[ \partial _{\alpha }\left( \alpha
\partial _{\alpha }B\right) +\alpha \partial _{\beta }\partial _{\beta }B%
\right] =2\partial _{u}\varphi \partial _{v}\varphi  \label{(uv)eq2bis}
\end{equation}%
with $\widetilde{K}_{up}^{p}=\partial _{u}\ln \alpha $. Writing out the $%
\left( uv\right) =\left( \alpha \beta \right) $ component directly returns (%
\ref{(uv)eq4 2}). The $\left( uv\right) =\left( \alpha \alpha \right) $ and $%
\left( uv\right) =\left( \beta \beta \right) $ components read 
\begin{subequations}
\begin{eqnarray}
2\partial _{\alpha }B-\partial _{\alpha }\left( \alpha \partial _{\alpha
}B\right) -\alpha \partial _{\beta }\partial _{\beta }B &=&2\alpha \partial
_{\alpha }\varphi \partial _{\alpha }\varphi  \label{(alpha-alpha)eq} \\
\partial _{\alpha }\left( \alpha \partial _{\alpha }B\right) +\alpha
\partial _{\beta }\partial _{\beta }B &=&-2\alpha \partial _{\beta }\varphi
\partial _{\beta }\varphi .  \label{(beta-beta)eq}
\end{eqnarray}%
Summing (\ref{(alpha-alpha)eq}) and (\ref{(beta-beta)eq}) returns (\ref%
{(uv)eq4 1}). Inserting $\partial B$\ from (\ref{(uv)eq4 1})-(\ref{(uv)eq4 2}%
), the remaining equation, for instance (\ref{(beta-beta)eq}), is satisfied
thanks to (\ref{KG4}).

\bigskip

\noindent \textbf{References}

\noindent \lbrack 1] A. Krasi\'{n}ski, \textit{Inhomogeneous Cosmological
Models}, Cambridge University Press (1997).

\noindent \lbrack 2] J. B. Griffiths, J. Podolsk\'{y}, \textit{Exact
Space-Times in Einstein's General Relativity}, (Cambridge University Press,
2009).

\noindent \lbrack 3] K.\ Bolejko, A. Krasi\'{n}ski, C.\ Hellaby, M.-N. C\'{e}%
l\'{e}rier, \textit{Structures in the Universe by exact methods} (Cambridge
University Press, 2010).

\noindent \lbrack 4] H.\ Stephani, D.\ Kramer, M.\ MacCallum, C.\
Hoenselaers, E.\ Herlt, \textit{Exact solutions of Einstein's field equations%
} (Cambridge University Press, 2003).

\noindent \lbrack 5] Y.\ Fujii, K. Maeda, \textit{The scalar-tensor theory
of gravitation} (Cambridge University Press, 2003).

\noindent \lbrack 6] V.\ Faraoni, \textit{Cosmology in scalar-tensor gravity}
(Kluwer Academic Publishers, 2004).

\noindent \lbrack 7] S.\ Capozziello, V. Faraoni, \textit{Beyond Einstein's
gravity}, Fundamental Theories of Physics, volume 170, Springer (2011).

\noindent \lbrack 8] C.\ M.\ Will, \textit{The confrontation between general
relativity and experiments }in www.livingreviews.org/Irr-2014-4 (living
reviews in relativity).

\noindent \lbrack 9] T.\ Damour, K.\ Nordtvedt, Phys.\ Rev.\ Lett. \textbf{70%
}, 2217 (1993).

\noindent \lbrack 10] T.\ Damour, K.\ Nordtvedt, Phys.\ Rev.\ D \textbf{48},
3436 (1993).

\noindent \lbrack 11] C. Brans and R. H. Dicke, Phys. Rev. \textbf{124}, 925
(1961).

\noindent \lbrack 12] C. H. Brans, Phys. Rev. \textbf{125}, 2194 (1962).

\noindent \lbrack 13] V.\ Faraoni, F. Hammad, S.\ D.\ Belknap-Keet, Phys.
Rev. D \textbf{94}, 104019 (2016).

\noindent \lbrack 14]\ K.\ A.\ Bronnikov, Acta Phys. Pol. \textbf{B}4, 251
(1973).

\noindent \lbrack 15] B.\ Chauvineau, Gen. Rel. Grav. \textbf{49}, 143
(2017).

\noindent \lbrack 16] S. W. Hawking, Commun. Math. Phys. \textbf{25}, 167
(1972).

\noindent \lbrack 17] J. D. Bekenstein and A. Meisels, Phys. Rev. D \textbf{%
18}, 4378 (1978).

\noindent \lbrack 18] R. H. Dicke, Phys. Rev. \textbf{125}, 2163 (1962).

\noindent \lbrack 19] C. B. G. McIntosh, Commun.\ Math. Phys. \textbf{37},
335 (1974).

\noindent \lbrack 20] R.\ N.\ Tiwari, B.\ K.\ Nayak, Phys. Rev. D \textbf{14}%
, 2502 (1976).

\noindent \lbrack 21] B.\ K.\ Nayak, R.\ N.\ Tiwari, J.\ Math. Phys. \textbf{%
18}, 289 (1977).

\noindent \lbrack 22] T.\ Singh, L.\ N.\ Rai, J. Math. Phys. \textbf{22},
136 (1981).

\noindent \lbrack 23] T.\ Singh, T.\ Singh, Astrophys. Sp. Sc. \textbf{100},
309 (1984).

\noindent \lbrack 24] A.\ Garc\'{\i}a N.\ Bret\'{o}n, I.\ Hauser, Int. J.
Theor. Phys. \textbf{27}, 635 (1988).

\noindent \lbrack 25] G.-W. Ma, Int. J. Theor. Phys. \textbf{34}, 2331
(1995).

\noindent \lbrack 26] H. Kim, Phys. Rev. D \textbf{60}, 024001 (1999).

\noindent \lbrack 27] J.\ Sultana, B.\ Bose, Phys. Rev. D \textbf{92},
104022 (2015).

\noindent \lbrack 28] P.\ Kirezli, \"{O}. Delice, Phys. Rev. D \textbf{92},
104045 (2015).

\noindent \lbrack 29] M.\ Sharif, R.\ Manzoor, Commun. Theor. Phys. \textbf{%
68}, 39 (2017).

\noindent \lbrack 30] V.\ Faraoni, D.\ K.\ \c{C}iftci, S.\ D.\ Belknap-Keet,
Phys. Rev. D \textbf{97}, 064004 (2018).

\noindent \lbrack 31] K.\ D.\ Krori, D. R. Bhattacharjee, J. Math. Phys. 
\textbf{23}, 637 (1982).

\noindent \lbrack 32] K. K. Nandi, P.\ M.\ Alsing, J.\ C.\ Evans, T.\ B.\
Nayak, Phys. Rev. D \textbf{63}, 084027 (2001).

\noindent \lbrack 33] G. N. Gyulchev, S. S. Yazadjiev, Phys. Rev. D \textbf{%
78}, 083004 (2008).

\noindent \lbrack 34] Z. Kov\'{a}cs, T. Harko, Phys. Rev. D \textbf{82},
124047 (2010).

\noindent \lbrack 35] T. Karmakar, T.\ Sarkar, Gen. Rel. Grav. \textbf{50},
85 (2018).

\noindent \lbrack 36] M.\ A.\ Makukov, E.\ G.\ Mychelkin, Phys. Rev. D 
\textbf{98}, 064050 (2018).

\noindent \lbrack 37] B.\ Chauvineau, Phys. Rev. D \textbf{98}, 088501
(2018).

\noindent \lbrack 38] E. T. Newman, A. I. Janis, J. Math. Phys. \textbf{6},
915 (1965).

\noindent \lbrack 39] H.\ Erbin, Gen. Rel. Grav. \textbf{47}, 19 (2015).

\noindent \lbrack 40] H.\ Erbin, Universe \textbf{3}, 19 (2017).

\noindent \lbrack 41] G. Cl\'{e}ment, Phys. Rev. D \textbf{57}, 4885 (1998).

\noindent \lbrack 42] H.\ Levy, W.\ J.\ Robinson, Proc. Camb. Phil. Soc. 
\textbf{60}, 279 (1964).

\noindent \lbrack 43] Z. Kov\'{a}cs, T. Harko, S.\ Shahidi, Phys. Rev. D 
\textbf{98}, 088502 (2018).

\noindent \lbrack 44] A. I. Janis, E. T. Newman, J. Winicour, Phys. Rev.
Lett. \textbf{20}, 878 (1968).

\noindent \lbrack 45] I.\ Z.\ Fisher, Z. Exp. Teor. Fiz. \textbf{18}, 636
(1948); see also arXiv:gr-qc/9911008 for an English translation.

\noindent \lbrack 46] R. Penney, Phys. Rev. \textbf{174}, 1578 (1968).

\noindent \lbrack 47] B.\ Turimov, B.\ Ahmedov, M.\ Kolo\v{s}, Z.\ Stuchl%
\'{\i}k, arXiv:gr-qc/1810.01460 (2018).

\noindent \lbrack 48]\ R. V. Wagoner, Phys. Rev. D \textbf{1}, 3209 (1970).

\noindent \lbrack 49] B.\ Chauvineau, Gen. Rel. Grav. \textbf{39}, 297
(2007).
\end{subequations}

\end{document}